# Inverse Design of Optical Multilayer Thin Films using Robust Masked Diffusion Models

Jonas Schaible[1,2,*], Asena Karolin Özdemir[3], Charlotte Debus[3], Sven Burger[2,4], Achim Streit[3], Christiane Becker[1,5], Klaus Jäger[1,2], and Markus Götz[3,6,+]

[1]Department Optics for Solar Energy, Helmholtz-Zentrum für Materialien und Energie GmbH, Berlin, Germany
[2]Zuse Institute Berlin, Berlin, Germany
[3]Scientific Computing Center, Karlsruhe Institute of Technology, Eggenstein-Leopoldshafen, Germany
[4]JCMwave GmbH, Berlin, Germany
[5]Hochschule für Technik und Wirtschaft Berlin, Berlin, Germany
[6]Helmholtz AI, Karlsruhe, Germany
[*]e-mail: jonas.schaible@helmholtz-berlin.de
[+]e-mail: markus.goetz@kit.edu

## ABSTRACT

Inverse design of optical multilayer stacks seeks to infer layer materials, thicknesses, and ordering from a desired target spectrum. It is a long-standing challenge due to the large design space and non-unique solutions. We introduce `OptoLlama`, a masked diffusion language model for inverse thin-film design from optical spectra. Representing multilayer stacks as sequences of material-thickness tokens, `OptoLlama` conditions generation on reflectance, absorptance, and transmittance spectra and learns a probabilistic mapping from optical response to structure. Evaluated on a representative test set of 3,000 targets, `OptoLlama` reduces the mean absolute spectral error by 2.9-fold relative to a nearest-neighbor template baseline and by 3.45-fold relative to the state-of-the-art data-driven baseline, called `OptoGPT`. Case studies on designed and expert-defined targets show that the model reproduces characteristic spectral features and recovers physically meaningful stack motifs, including distributed Bragg reflectors. These results establish diffusion-based sequence modeling as a powerful framework for inverse photonic design.

## 1 Main

OPTICAL multilayer thin films consist of multiple wavelength-scale layers of different materials, each characterized by distinct optical properties. By carefully controlling the interaction of light with these layer stacks, for example through material choice, stacking order, and layer thickness, it is possible to realize desired spectral responses in reflectance, absorptance, and transmittance[1]. As a result, optical multilayer thin films are widely used across diverse application domains, including optical filters such as color filters[2,3], bandpass filters[4], and bandstop or spectrally selective filters[5], photovoltaic solar cells[6], absorptive devices such as perfect absorbers[7], transparent electrodes[8,9], and reflective structures for radiative cooling[10–12].

Traditionally, multilayer thin films have been designed using a forward approach, in which material choices and layer configurations are specified a priori and the resulting optical properties are evaluated through simulation or experimental characterization. Desired performance is then achieved through iterative refinement of the material choice, layer sequence, and thicknesses. In practice, this procedure is time-consuming, error-prone, and costly because of the combinatorially large design space[13,14].

In contrast, inverse design formulates the problem by specifying optical target properties and computationally inferring suitable material combinations, layer arrangements, and thicknesses[15].

Inverse design of optical multilayer stacks has been studied for decades, traditionally relying on gradient-based optimization, needle-variation methods, and evolutionary algorithms to iteratively match target spectra through repeated forward simulations[15,16]. Although these optimization-based methods have enabled high-performance coatings, they are often computationally expensive and provide limited insight into the non-uniqueness of valid solutions.

Recent machine learning-based approaches learn data-driven inverse mappings that enable faster inference and the generation of multiple candidate designs. In inverse photonic design, such methods have been applied to structures including metasurfaces and multilayers[14,17,18]. For optical multilayer thin films, recent work has increasingly focused on learning the inverse mapping from target spectra to corresponding layer stacks[19]. Autoregressive approaches such as `OptoGPT`[20] have advanced the state of the art by enabling spectra-conditional generation of variable-depth stacks and stochastic sampling of alternative solutions. Nevertheless, important challenges remain, including limited predictive fidelity, restricted thickness representation, and constrained design diversity arising from model overconfidence.

From a modeling perspective, a multilayer stack can be represented as an ordered sequence of discrete tokens, analogous to words in a sentence, with each "word" or token corresponding to a material-thickness pair (cf. Figure 1a). This formulation casts inverse thin-film design as a sequence

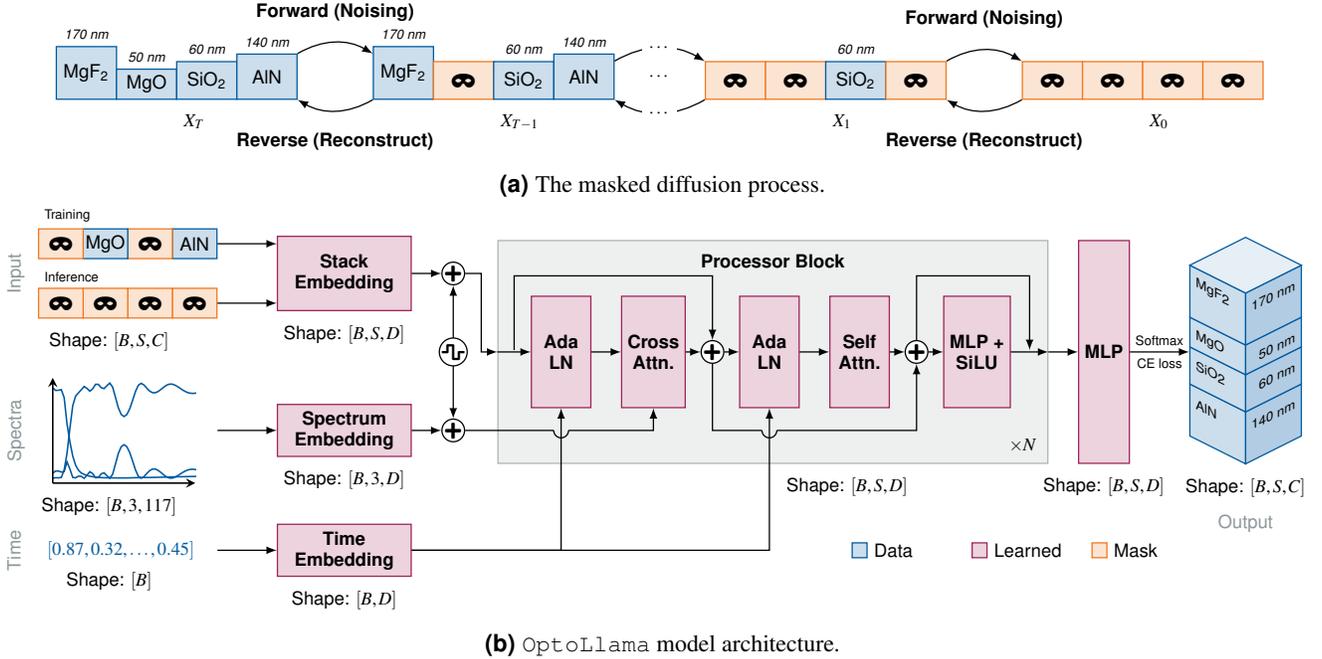

**(a)** The masked diffusion process.

**(b)** `OptoLlama` model architecture.

**Figure 1. `OptoLlama` framework.** Overview of the `OptoLlama` framework for inverse thin-film design. The masked diffusion process is illustrated through its forward (noising) and reverse (reconstruction) stages (Figure 1a), together with a high-level view of the model architecture and its input–output interfaces (Figure 1b).

modeling problem and enables the use of conditional generative architectures from natural language processing[21].

In this work, we introduce `OptoLlama`, a masked diffusion framework for inverse multilayer thin-film design. We condition generation on reflectance, absorptance, and transmittance spectra as separate channels and, to our knowledge, provide the first application of a masked diffusion language model to inverse thin-film design. Unlike autoregressive approaches, the diffusion-based formulation considers the full layer stack sequence simultaneously, enabling globally consistent design decisions. The model starts from a fully masked sequence and iteratively replaces mask tokens with material-thickness tokens through a learned denoising procedure (cf. Figures 1a and 1b). Using this approach, `OptoLlama` reduces the mean absolute spectral error by 2.9-fold relative to a nearest-neighbor template baseline and by 3.45-fold relative to a reproduced `OptoGPT` baseline.

The model generates layer stacks that reflect established design principles, including distributed Bragg reflector (DBR) motifs, that is, alternating high- and low-index multilayer substructures that produce strong reflection in a target wavelength range. Detailed statistics on DBR-like motif occurrence are provided in the Supplementary Information (Section 1.13). In addition to achieving low mean absolute errors compared to the target spectra, the generated designs preserve characteristic spectral features, including bandstop filter and color-selective peaks. Design flexibility is addressed via data-centric conditioning and model-driven Monte Carlo sampling enabled by the probabilistic nature of the model. In contrast to autoregressive approaches, where diversity arises primarily from stochastic decoding, the diffusion-based formulation explicitly learns a conditional probability distribution over material stacks, from which samples are drawn at each denoising step. The incorporation of practical constraints, such as limits on layer number or allowed materials and thicknesses at specific positions reflect fabrication or operator requirements. Finally, we provide an open implementation including curated training and evaluation datasets, source code, and pretrained model checkpoints.

## 2 Results

`OptoLlama` formulates inverse thin-film design as a conditional generative problem, in which target reflectance, absorptance, and transmittance (RAT) spectra are provided as conditioning input and candidate multilayer stacks are generated through a masked diffusion process (cf. Figure 1). In this setting, a multilayer stack is represented as an ordered sequence of thin-film layers, and each token corresponds to a material-thickness combination drawn from a finite vocabulary. Starting from a fully masked token sequence, the model iteratively denoises the stack by replacing mask tokens with plausible tokens conditioned on both the partially reconstructed sequence and the target spectra. At each reversal step, the model predicts a conditional distribution over tokens, from which samples are drawn to progressively refine the sequence. Because this reverse process is stochastic, `OptoLlama` can generate multiple candidate solutions for

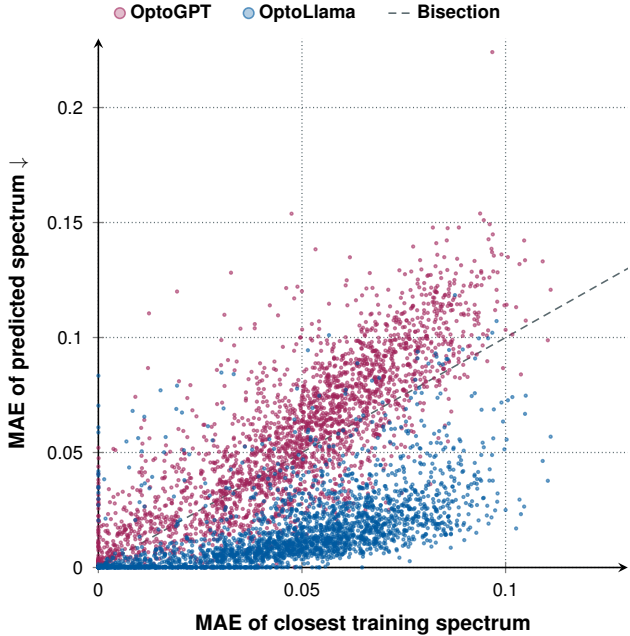 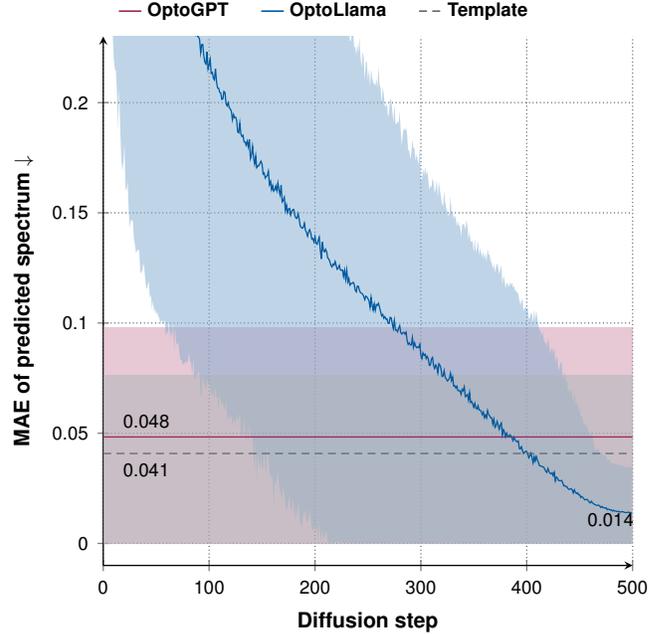

**(a)** Predictive performance relative to template baseline.   **(b)** Prediction quality across diffusion steps.

**Figure 2. Aggregated performance.** (a) Mean absolute error (MAE) of `OptoLlama` predictions compared with the template baseline on 3,000 target spectra. (b) Mean MAE trajectory of `OptoLlama` across denoising steps, shown together with the reproduced `OptoGPT` and template baseline. Shaded bands indicate the 10th to 90th percentile range across targets.

the same spectral target. We leverage this property throughout the evaluation using Monte Carlo (MC) sampling.

An overview of the model architecture is shown in Figure 1b. At a high level, the model is a transformer with alternating self-attention blocks over stack tokens and cross-attention layers that correlate the stack representation with the target spectra.

### 2.1 Predictive performance

`OptoLlama` is evaluated on a representative test subset of 3,000 target spectra. Figure 2a compares the mean absolute error (MAE) from the target for stacks predicted by `OptoLlama` with that of the closest matching spectrum in the training set, referred to as the template baseline. Lower MAE indicates a closer match between the spectrum of the predicted stack and the target spectrum. For each target, the lowest MAE among ten Monte Carlo samples is reported for both `OptoLlama` and the autoregressive baseline `OptoGPT`[20], which we adapt to the present input dimensionality and train for 1,000 epochs.

For `OptoGPT`, most points lie close to the bisection line, indicating performance similar to the template baseline. By contrast, the majority of points for `OptoLlama` lie below the bisection line, showing systematic improvement over both pure lookup-based design and the reproduced autoregressive baseline. Most targets are matched with low MAE values, whereas only a small fraction exhibit higher errors than the template baseline. Overall, these results show that `OptoLlama` achieves consistently lower reconstruction errors than template baseline designs, providing strong empirical evidence of generalization beyond the training set.

Figure 2b compares `OptoLlama` (blue), `OptoGPT` (red), and the template baseline (grey) on the same test subset of 3,000 target spectra. The plot reports the mean MAE together with the $10^{th}$ to $90^{th}$ percentile range as a function of diffusion step for `OptoLlama`, while `OptoGPT` and the template baseline appear as step-independent reference levels. Unlike in natural language processing, prediction quality can be quantified at each diffusion step by simulating the spectra of the intermediate predicted stacks, yielding an MAE trajectory for every target. Although the percentile bands indicate substantial variability across targets, the overall trend remains consistent.

`OptoLlama` exhibits higher errors during the early denoising stages, particularly during the first $\approx 400$ diffusion steps, but its MAE decreases steadily as the sequence is refined and ultimately surpasses both baselines. At the final diffusion step, `OptoLlama` reaches a mean MAE of 0.014, compared to 0.041 for the template baseline and 0.048 for the reproduced `OptoGPT` model, corresponding to 2.9-fold and 3.45-fold lower error, respectively. Interestingly, `OptoLlama` underperforms both baselines for roughly the first four-fifths of the denoising trajectory and only matches and then surpasses them late in the reverse process. This behavior indicates that high-fidelity stack reconstruction emerges only late in the reverse process after substantial iterative refinement.

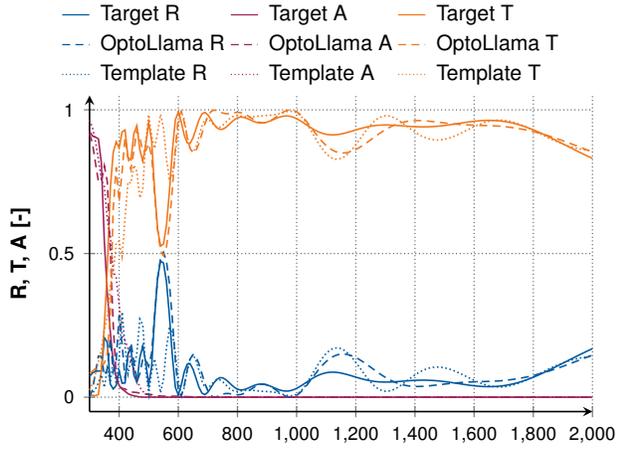

**(a)** Spectrum for color filter.

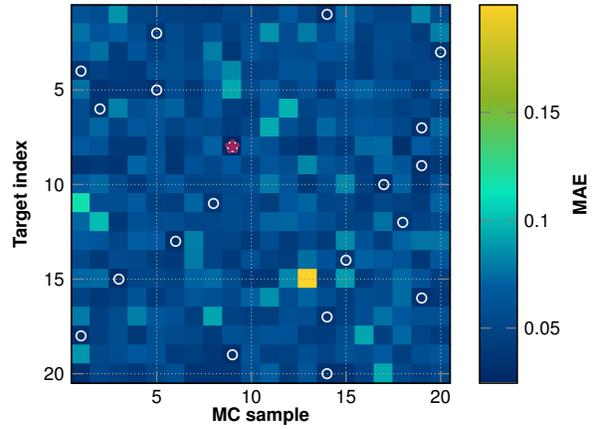

**(b)** Monte Carlo (MC) sampling for color filter.

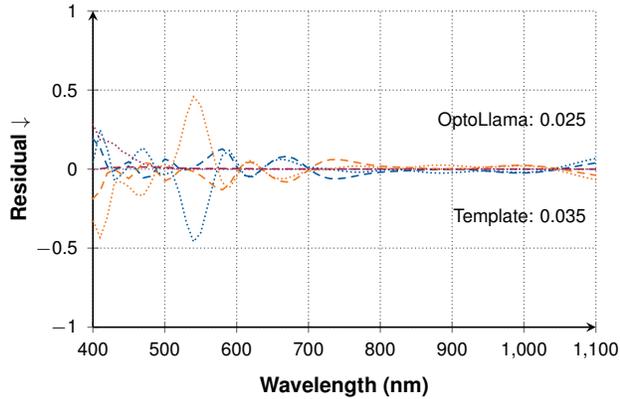

**(c)** residuals for color filter.

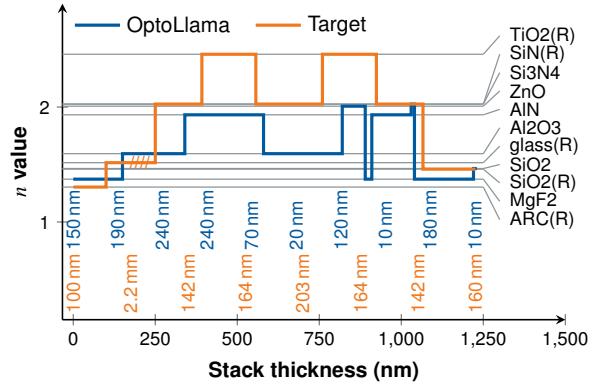

**(d)** Refractive-index profile of the color filter at 550 nm.

**Figure 3. Example inverse design: MorphoColor-inspired color filter.** Four panels summarize the target spectrum, stochastic generation process, and best-performing design produced by `OptoLlama`. The top row shows the target and reconstructed reflectance, absorptance, and transmittance spectra (left) and an overview of MAE values across Monte Carlo samples (right), with row-wise best samples and the global best highlighted. The bottom row shows wavelength-resolved residuals (left) and the refractive-index profile of the selected design at 550 nm (right). For comparison, the bottom-right panel also includes the expert-designed target stack.

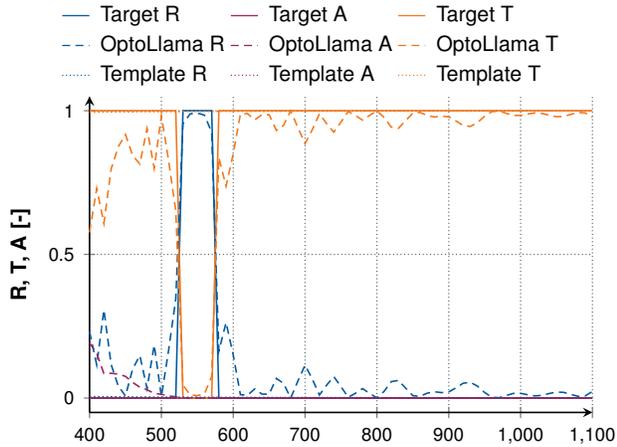

**(a)** Spectrum for bandstop filter.

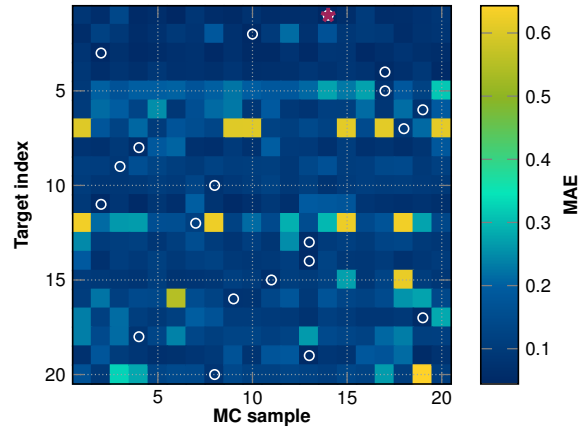

**(b)** Monte Carlo (MC) sampling for bandstop filter.

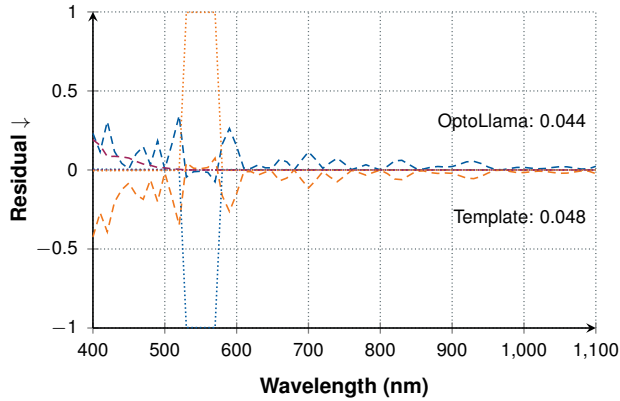

**(c)** residuals for bandstop filter.

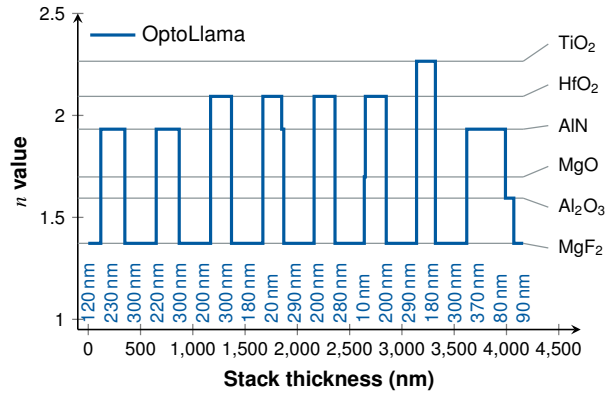

**(d)** Refractive-index profile of the bandstop filter at 550 nm.

**Figure 4. Example inverse design: artificial bandstop target.** Four panels summarize the target spectrum, stochastic runs, and best-performing design produced by `OptoLlama`. The top row shows the target and reconstructed reflectance, absorptance, and transmittance spectra (left) and an overview of MAE values across Monte Carlo samples (right), with row-wise best samples and the global best highlighted. The bottom row shows wavelength-resolved residuals (left) and the refractive-index profile of the selected design at 550 nm (right).

## 2.2 Case studies

We evaluate `OptoLlama` on two representative inverse-design targets shown in Figures 3 and 4: a realistic, expert-designed MorphoColor-inspired color filter and an artificially constructed bandstop target. This analysis complements the aggregated benchmark by assessing reconstruction accuracy, feature preservation, and physical interpretability at the individual-design level.

For each target, 20 independent noise initializations are combined with 20 repeated stochastic evaluations, yielding a $20 \times 20$ set of Monte Carlo samples. Unless noted otherwise, the spectral comparison, residual plot, and refractive-index profile shown for each example correspond to the globally best-performing realization across these samples.

**Color filter — real-world target.** Figure 3 considers a MorphoColor®-inspired color filter target[22], which has been experimentally demonstrated in previous work[3]. This concept enables solar modules to be tuned to a range of target colors; here we focus on a representative green design. Because the target spectrum is based on a real MorphoColor®sample, this example provides an experimentally grounded application of whether `OptoLlama` can reproduce an established thin-film design. A comparison between experiment and simulation is provided in the Supplementary Information (Figure S3).

Figure 3a shows the target as solid lines, the `OptoLlama` prediction as dashed lines, and the template baseline as dotted lines. `OptoLlama` reproduces the characteristic spectral response with high fidelity (MAE = 0.025), accurately capturing both the position and amplitude of the reflectance peak near 550 nm. By contrast, the template baseline exhibits larger residuals (MAE = 0.0350) and fails to reproduce the defining spectral feature.

The corresponding Monte Carlo MAE matrix is shown in Figure 3b. Row-wise minima are indicated by circles and the global minimum by a star, highlighting the variability across stochastic realizations and the best-performing sampled design.

Consistently, the residuals of `OptoLlama` are smaller than those of the template baseline (Figure 3c).

The refractive-index profile of the predicted stack, shown together with the expert-designed target stack in Figure 3d, indicates that both designs begin with low-index entrance layers and span a comparable overall thickness range. However, the predicted profile does not reproduce the target stack motif directly and does not exhibit an immediately obvious established design motif. This suggests that `OptoLlama` recovers a different but physically valid interference strategy within its discrete solution space, consistent with the possibility that inverse design can identify non-obvious structural design rules in photonic systems[23].

**Bandstop filter — artificial target.** Figure 4 presents an artificially constructed bandstop target centered at 550 nm with a width of 50 nm. For comparability with prior work using `OptoGPT`[20], the target is specified over 400 nm to 1100 nm, while the remaining wavelengths are filled with randomly generated smoothed values to retain a full-spectrum target.

Figure 4a shows the target as solid lines, the `OptoLlama` prediction as dashed lines, and the template baseline as dotted lines. Despite the unrealistic box-like spectral shape, `OptoLlama` closely follows the desired response (MAE = 0.044) and reproduces the characteristic stop-band behavior. The template baseline from the training set exhibits slightly larger residuals on average (MAE = 0.048) and does not reproduce the bandstop feature.

The corresponding Monte Carlo MAE matrix is shown in Figure 4b. The residuals are shown in Figure 4c.

The refractive-index profile in Figure 4d exhibits a pronounced alternating high/low-index structure between $MgF_2$ and higher-index materials including AlN, $HfO_2$, and $TiO_2$, reminiscent of DBR-like substructures. The corresponding layer pairs are consistent with third-order quarter-wave conditions at approximately 564 nm, 548 nm, and 542 nm, all within the target bandstop window around 550 nm, whereas the corresponding first-order peaks fall near 1600 nm to 1700 nm. Overall, the predicted stack is consistent with the design logic of a third-order quarter-wave reflector; detailed wavelength estimates are provided in the Supplementary Information (Table S2). In contrast, the template baseline stack is composed solely of a $MgF_2$ layer of 10 nm.

To test whether these DBR-like motifs reflect memorization, we quantified their occurrence in both the training data and Monte Carlo samples. Among the 10 million training stacks, 932,237 (9.3%) contained at least one DBR motif, almost always of length 3 and only rarely of length 5. By contrast, DBR motifs were strongly enriched in `OptoLlama` outputs, especially for the bandstop target, for which all 400 samples contained at least one DBR and longer chains of length 7, 9, and 11 were observed. These results suggest that the model does not simply reproduce training examples, but preferentially assembles DBR-like structures under task-specific spectral conditioning.

Additional inverse-design examples, including a visually transparent thermal reflector and a perfect absorber, are provided in the Supplementary Information (Figures S6 and S7).

## 2.3 Variability analysis

We next analyze the variability of `OptoLlama`'s outputs across Monte Carlo samples for the bandstop target and the MorphoColor-inspired color filter target.

In both cases, the first layer is predicted as magnesium fluoride ($MgF_2$) with a probability exceeding 84% (see Figures 5a and 5b). This preference is physically plausible because both targets exhibit zero or near-zero reflection across large parts of the spectrum, favoring a low-index entrance layer that minimizes front-surface reflection through a smooth optical transition from air into the multilayer stack. Among the available materials, $MgF_2$ has the lowest refractive index (Figure S1).

In both investigated cases, the second layer is most fre-quently predicted as either $SiO_2$ or MgO, corresponding to materials with similarly low refractive index. Together, these first layers create a graded-index-like entrance region that suppresses reflection over a broad spectral range. From the third layer onward, the bandstop and color filter examples begin to diverge, although both require a localized reflection feature within a specific wavelength range. In classical thin-film optics, such features are commonly realized using DBRs[1], in which alternating high- and low-index layers are arranged with approximately quarter-wave optical thickness. Consistent with this design principle, OptoLlama frequently selects low-index dielectrics such as $MgF_2$, $SiO_2$, and MgO, together with higher-index materials such as $Al_2O_3$, AlN, and $HfO_2$, to build the stack (Figure S1). These variability patterns indicate that the model does not merely memorize individual layer sequences, but rather captures families of optically similar materials that can be interchanged while preserving the desired spectral response. Additional variability heatmaps for the thermal-reflector and perfect-absorber targets are shown in the Supplementary Information (Figure S8).

### 2.4 Energy usage

In line with recent efforts toward more sustainable AI research[24], we report the energy consumption associated with a single training run, 100 inference runs, and the overall development process. The resulting energy-use estimations, $CO_2$-equivalent emissions, and monetary cost estimates are summarized in Table 1. All energy estimations were obtained by scaling the average quantities reported for a transformer training runs as reported in prior studies[25]. Details of the accounting procedure, including the formulas used to estimate $CO_2$-equivalent emissions and cost, the assumed power usage effectiveness (PUE), and the data sources for carbon intensity and electricity price, are provided in the Supplementary Information (Section 1.8).

## 3 Discussion

The results demonstrate that OptoLlama can accurately perform inverse design of optical multilayer thin films across both realistic and artificially constructed targets, while reproducing characteristic spectral features and physically meaningful material-stack patterns. These capabilities are achieved despite several simplifying assumptions in the optical simulation model and dataset construction, which we discuss below.

The optical simulation backbone of OptoLlama is currently based on a transfer-matrix-method (TMM) solver[26,27] operating under normal incidence and assuming laterally homogeneous, planar layers. Effects such as polarization dependence, oblique incidence, surface roughness, incoherent propagation of light in optically thick layers, and lateral patterning are not considered. Although these simplifications limit direct applicability under some experimental conditions, they enable fast and physically consistent simulations that capture the dominant interference effects governing multilayer optics. OptoLlama's ability to reproduce physically interpretable multilayer motifs, including DBRs, suggests that the model learns meaningful thin-film design principles rather than merely overfitting to artifacts of the simulation data. Whereas the bandstop example admits a relatively direct interpretation in terms of DBR-like substructures, the color filter example suggests that high-fidelity spectra can also arise from less established design motifs and other interference strategies.

The training dataset consists of exact forward simulations and does not explicitly account for fabrication tolerances such as layer-thickness variations or uncertainty in refractive indices. Nevertheless, the probabilistic nature of diffusion-based generation partially mitigates this limitation: repeated stochastic runs produce multiple alternative material stacks with comparable spectral performance that can subsequently be screened for robustness. The variability analysis shows diminishing gains beyond a moderate number of Monte Carlo samples, suggesting that near-optimal solutions can be obtained efficiently while still allowing exploration of the design space.

A further limitation arises from the discrete tokenization of materials and thicknesses. Currently, layer thicknesses are restricted to multiples of 10 nm, and the wavelength grid is similarly discretized. As a consequence, designs requiring continuous or finer-grained parameter choices are excluded from the representable space. Despite this discretization, OptoLlama consistently matches target spectra with low mean absolute error and preserves key spectral characteristics, indicating that the inverse-design problem is already well approximated within this discrete representation. Increasing the resolution of the token space is conceptually straightforward, but would require a substantially larger training dataset to adequately cover the expanded combinatorial space.

The material palette is likewise limited to a predefined set of metals, semiconductors, and dielectrics. Although this restricts extrapolation to other materials, it ensures physical consistency and reliable optical properties throughout training and inference. Incorporating additional materials would require retraining or fine-tuning the model with updated optical constants, but the underlying framework is agnostic to the specific material set and can be extended readily.

A central finding of this work is that the masked diffusion approach provides substantial advantages over autoregressive approaches for inverse multilayer design. By operating on the entire token sequence in parallel, the diffusion process reduces the error accumulation and overconfidence associated with autoregressive generation. This is reflected in the improved predictive fidelity, reduced MAE, and increased diversity of generated stacks compared with OptoGPT. Moreover, the explicit probabilistic formulation enables principled uncertainty quantification through Monte Carlo sampling, which is not naturally available in purely autoregressive token sampling schemes.

Although OptoLlama has slightly more parameters than the reproduced OptoGPT baseline (111 555 513 vs.

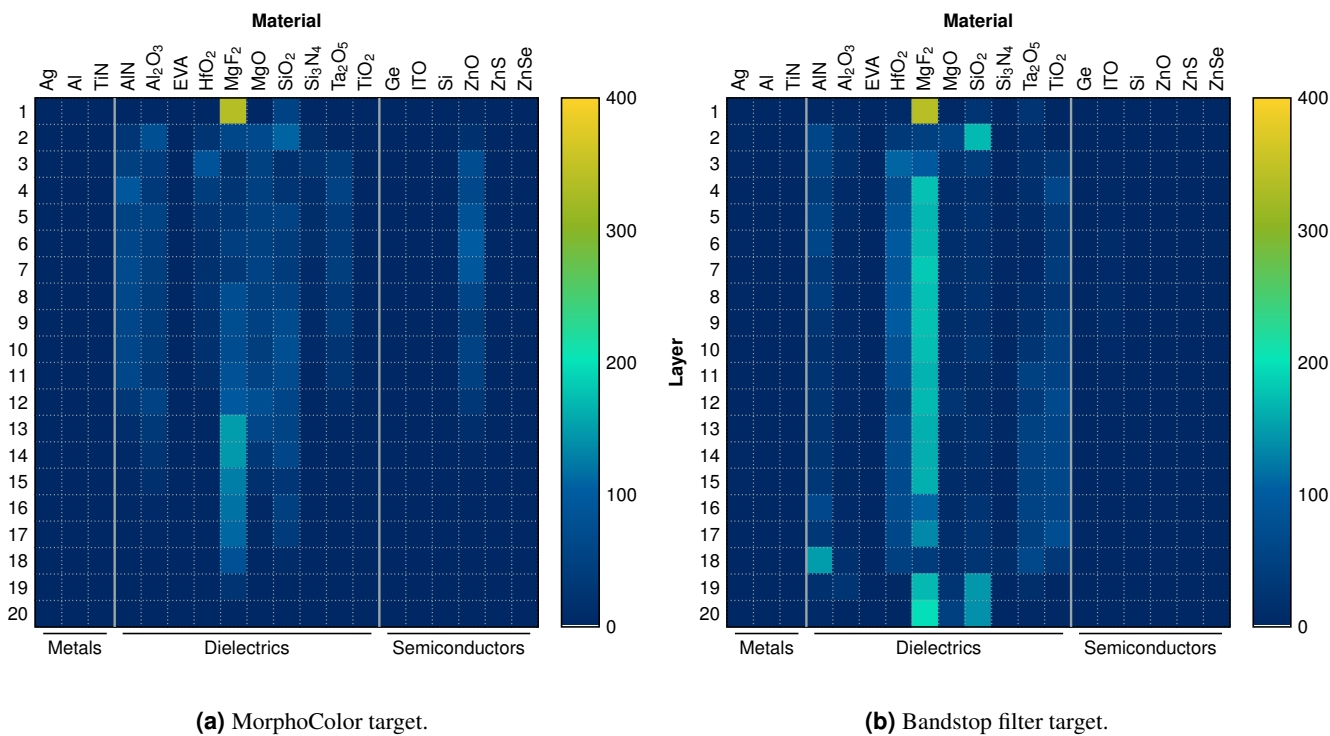

**(a)** MorphoColor target.  **(b)** Bandstop filter target.

**Figure 5. Material variability heatmaps across Monte Carlo samples.** For each target task (a,b), we run `OptoLlama` with $N = 400$ Monte Carlo samples and record the decoded material token at each layer position. Each heatmap shows the resulting material count per thin-film layer (rows) and material class (columns); brighter colors indicate higher frequency. Horizontal brackets group materials into metals, dielectrics, and semiconductors. These distributions summarize model variability and highlight materials that are frequently interchangeable at specific layer positions, as discussed in Variability analysis.

**Table 1.** Energy usage estimation, carbon emission equivalents and monetary cost of the `OptoLlama` model for training, inference and overall development. Calculation details are provided in the Supplementary Information Section 1.8.

| Experiment and location | Energy usage [kWh] | Carbon efficiency [kg/MWh] | $CO_2$ equivs. [kg] | Cost efficiency [$/kWh] | Energy cost [$] |
|---|---|---|---|---|---|
| **Model training** | | | | | |
| California, USA | | 340.0 | 273.03 | 0.32 | 256.97 |
| France | 803.04 | 23.0 | 18.46 | 0.28 | 224.85 |
| China | | 553.0 | 444.08 | 0.08 | 64.24 |
| **Inference (100 samples)** | | | | | |
| California, USA | | 340.0 | 0.00340 | 0.32 | 0.0032 |
| France | 0.01 | 23.0 | 0.00023 | 0.28 | 0.0028 |
| China | | 553.0 | 0.00550 | 0.08 | 0.0008 |
| **Total development** | | | | | |
| California, USA | | 340.0 | 9,669.94 | 0.32 | 9,101.12 |
| France | 28,441.00 | 23.0 | 654.14 | 0.28 | 7,963.48 |
| China | | 553.0 | 15,727.87 | 0.08 | 2,275.28 |

108 381 113), this corresponds to only a 2.9% increase in model size. Such a marginal increase is unlikely to explain the substantially larger MAE reduction. Consistent with neural scaling laws, we therefore attribute the performance gains primarily to the masked diffusion formulation and conditioning strategy rather than to parameter count alone[28, 29].

Precise, point-wise loss functions such as cross-entropy, as used in this work, are fundamentally ill-suited to this inverse-design setting because they penalize any deviation from a single "ground truth" design provided in the training data. If the model were to predict a secondary design, which is physically valid but different from the labeled primary design, a standard loss function treats this as an error and biases the model toward reproducing the labeled primary solution only. In other words, such deterministic losses fail to capture the multi-modal distribution of the design space. We encourage future studies to tackle this fundamental problem with a more suitable loss function design.

Finally, although reflectance, absorptance, and transmittance are physically linked by energy conservation, providing all three as separate input channels makes absorption-related spectral features more explicit to the model and therefore encourages solutions that are more sensitive to absorptive design goals. This is particularly relevant for applications such as photovoltaics and absorptive coatings, where small deviations in absorption can be critical. Physical consistency is nevertheless preserved by the forward solver, ensuring that generated designs obey fundamental conservation laws.

Overall, OptoLlama demonstrates that diffusion-based language models can serve as effective and flexible tools for solving ill-posed inverse problems in optics. By combining physically grounded simulation, probabilistic generation, and data-centric conditioning, the framework bridges deep learning and materials science and provides a foundation for future extensions toward more complex optical models, fabrication-aware datasets, and experimental closed-loop design.

This work introduces OptoLlama, a diffusion-based language-modeling framework for the inverse design of optical multilayer thin films. Building on a sequence-based representation of material stacks, OptoLlama establishes masked diffusion transformers as a powerful alternative to autoregressive approaches for solving ill-posed inverse problems in optics.

Compared with prior transformer-based methods such as OptoGPT[20] and template baseline-driven design, OptoLlama achieves substantially improved predictive fidelity and robustness. By generating entire layer sequences in parallel and explicitly modeling uncertainty through diffusion, the approach reduces error accumulation and overconfidence, leading to lower reconstruction errors and increased diversity of viable design solutions. These advantages are demonstrated quantitatively on a representative test set of 3,000 targets and qualitatively across representative synthetic and realistic design scenarios.

Although the present implementation relies on simplified optical simulations and a discrete material representation, the results indicate that diffusion-based inverse design can capture physically meaningful design principles even under these constraints. Future extensions toward more realistic optical models, expanded material libraries, and fabrication-aware datasets are natural next steps.

By releasing the full dataset, source code, and pretrained models, this work provides a shared foundation for further work at the interface of machine intelligence and materials science. More broadly, OptoLlama points toward a new class of probabilistic, physics-aware inverse-design methods in photonics and beyond.

## 4 Methods

### 4.1 Dataset generation

The dataset used in this work is based on that introduced by Ma et al.[20] and comprises 19 materials commonly used in optical multilayer thin films, including metals, semiconductors, and dielectrics. The wavelength-dependent complex refractive indices of all materials are provided in the Supplementary Information (Section 1.5). Each token sequence, corresponding to a multilayer material stack, is simulated using the transfer-matrix-method (TMM) implementation TMM-Fast[27]. All simulations assume planar, laterally homogeneous layers under normal incidence. Surface texture and incoherent effects are neglected, and light propagation is treated coherently throughout the stack. Spectra are computed over the wavelength range 300 nm to 2000 nm with a resolution of 10 nm, matching the spectral parameter space of OptoLlama. The distribution of training samples across material classes and stack positions is shown in the Supplementary Information (Figure S4).

**Tokenization and design space.** Tokens represent all combinations of the 19 materials with discrete thickness values ranging from 10 nm to 500 nm in steps of 10 nm, resulting in $19 \times 50 = 950$ unique layer tokens. Generated stacks are limited to a maximum of 20 layers, yielding a combinatorial design space of $950^{20} \approx 3.6 \times 10^{59}$ possible configurations. From this space, 11 million samples were generated in[20], of which 10 million are used for training and 1 million are reserved for testing.

**Optical simulation.** The optical response of each multilayer stack is computed using the transfer-matrix method under normal incidence. For a given material sequence and layer thicknesses, wavelength-dependent reflectance ($R$), transmittance ($T$), and absorptance ($A$) spectra are calculated. Complex refractive indices are interpolated linearly onto the simulation wavelength grid, and energy conservation is enforced by construction through the relation $A = 1 - R - T$. The same forward solver is used consistently for dataset generation, model evaluation, and validation of predicted designs.

### 4.2 Masked diffusion models

Diffusion models map a simple and tractable reference distribution to a complex data distribution $X$. During training, a clean data item $x_i$ in $X$ is corrupted by a forward process $q$ and subsequently reconstructed by a reverse process $p$. The degree of corruption is controlled by a parametric, monotonically increasing noising schedule $\alpha_t \in [0,1]$ and continuous time variable $t \in [0,1]$, such that $x_{i,t}$ denotes increasingly corrupted variants of $x_i$ as $t$ increases. Because a closed-form expression for the reverse process is generally intractable, it is approximated by a learned model $p_\theta$.

In classical diffusion modeling[30,31], the forward process gradually maps the data distribution to a standard Gaussian distribution. More recently, categorical diffusion models[32] have attracted increasing interest, particularly for applications in natural language processing[21,33]. In these models, parts of the input $x$, here tokens of the material sequence, are gradually replaced by a special mask token during the forward process. Conversely, the reverse process learns to restore the masked parts of the sequence. The learned model $p_\theta$ is trained to predict the corresponding clean input $x$.

We follow the formalism of masked diffusion language models (MDLM)[21]. The forward process defines a sequence of increasingly noisy latent variables $z_t$, with marginal

$$q(z_t|x_i) = \text{Cat}(z_t; \alpha_t x_i + (1-\alpha_t)\pi), \quad (1)$$

where each $x_i$ is a one-hot encoded vector over $K$ token classes, corresponding here to material-thickness combinations. Further, $\text{Cat}(\cdot;\pi)$ denotes a categorical distribution over $K$ classes with probabilities $\pi \in \Delta_K$, where $\Delta_K$ denotes the $K$-simplex. The $K$-th category corresponds to a special mask token $m$, i.e., $m_K = 1$. In masked diffusion, we set $\pi = m$, such that at each noising step $t$, the input transitions to the masked state with probability determined by $\alpha_t$. If an input transitions to $m$ at any time $t'$, it remains in this state for all $t > t'$, that is, $q(z_t|z_{t'}=m) = \text{Cat}(z_t;m)$. At time $t=1$, all inputs are masked with probability 1.

The MDLM framework uses a substitution-based parameterization of the reverse diffusion process:

$$p_\theta(z_s|z_t) = \begin{cases} \text{Cat}(z_s; z_t), & z_t \neq m, \\ \text{Cat}(z_s; \frac{(1-\alpha_s)m + (\alpha_s-\alpha_t)x_{i,\theta}(z_t,t)}{1-\alpha_t}), & z_t = m, \end{cases} \quad (2)$$

where $z_s$ denotes the latent variable at time $0 \leq s \leq t$. An unmasked token remains unchanged during reverse diffusion, whereas a masked token is iteratively restored. The training objective is the minimization of the token-level cross-entropy loss. Empirically, we find that the Rao-Blackwellized weighting proposed in MDLM [21], i.e., the introduction of an additional weighting term proportional to $t$, does not improve predictive performance and can in some cases degrade it. Detailed hyperparameters and the computational and software environments are provided in the Supplementary Information (Table S1 and sections 1.6 and 1.7).

**Spectral conditioning.** Model conditioning is performed using reflectance, absorptance, and transmittance spectra as separate input channels. In this context, the conditioning signal plays the role of a prompt, namely the target optical spectrum provided to guide generation of the multilayer stack. Although absorptance is deterministically related to reflectance and transmittance under energy conservation, explicitly including all three channels emphasizes absorption-related spectral features and introduces a favorable inductive bias into the learned inverse mapping. Physical consistency is ensured by the forward TMM solver, which enforces energy conservation for both training data and generated designs.

**Monte Carlo sampling and winner selection.** To assess robustness and variability, each target spectrum is evaluated using multiple Monte Carlo samples. Specifically, 20 independent noise initializations are combined with 20 repeated target realizations, yielding a $20 \times 20$ set of Monte Carlo samples per target. For each realization, the predicted stack is simulated using the forward TMM solver and evaluated using the mean absolute error with respect to the target spectrum. Row-wise minimum MAE values are used to assess per-run best performance, and the global minimum across all realizations is selected as the representative design shown in the detailed visualizations. Aggregated statistics across all Monte Carlo runs are used for the subsequent variability analysis.

### 4.3 Variability analysis approaches

Diffusion models are inherently probabilistic, which means that repeated evaluations of the same trained model on the same input can yield different outputs. However, a reliable method to determine how much meaningful difference actually exists between the results produced by different Monte Carlo samples of the trained model does not exist. This raises two practical questions for inverse design.

First, can Monte Carlo samples improve the final design quality if only the best-performing result is retained, and if so, how many stochastic runs are required before performance gains saturate? Because the predicted multilayer stacks can be evaluated with the forward optical solver, we compute the MAE for each individual model output. This analysis shows that different stochastic runs can indeed yield substantially different stacks and, correspondingly, different spectral errors. We therefore adopt a best-so-far strategy, in which the lowest-loss result observed across repeated stochastic runs is retained. Figure S9 shows how the best-so-far MAE over 3,000 targets improves as the number of stochastic runs increases. Although the MAE continues to decrease with additional runs, the rate of improvement slows markedly beyond approximately 10 runs. We therefore recommend at least 10 stochastic runs as a practical default that balances computational cost and solution quality, while noting that this number remains a task-dependent hyperparameter.

Second, can Monte Carlo samples reveal information about the interchangeability of materials within the generated stacks? More specifically, do certain groups of materials recur in similar layer positions across different runs, suggesting that the model has identified them as optically similar alternatives? To address this question, we analyze material-frequency heatmaps obtained from 400 Monte Carlo samples. These heatmaps (Figures 5a and 5b) reveal both strongly con-strained layer positions, where one material dominates, and less constrained positions, where several materials are used interchangeably. Comparing these patterns with known optical constants provides a physically grounded way to interpret model variability and to identify families of materials with similar functional roles in the generated designs.

### 4.4 Reflectance measurement

Experimental validation of selected thin-film designs is performed using an integrating sphere (Ulbricht sphere) to measure hemispherical reflectance. Samples are illuminated with broadband light over the wavelength range 300 nm to 2000 nm, and reflected intensities are collected over all outgoing angles. This configuration minimizes angular and polarization sensitivity. Measured spectra are corrected for the system response and referenced against calibrated standards. A comparison of measured and simulated reflectance for the MorphoColor®reference sample is shown in the Supplementary Information (Figure S3).

## Data availability

The dataset to reproduce the presented results is publicly available on HuggingFace and published under the Creative Commons Attribution 4.0 International Public License (CC-BY 4.0).

## Code availability

All code used in this work is publicly available on GitHub and is published under the Apache 2.0 license.

# Acknowledgements

This work is supported by the Helmholtz Association Initiative and Networking Fund through the Helmholtz International Berlin Research School in Data Science (HEIBRiDS), the Helmholtz AI platform, and the HAICORE@KIT grant. We gratefully acknowledge the computing time provided on the supercomputer Lise at NHR@ZIB as part of the NHR infrastructure. A part of the results were obtained at the Berlin Joint Lab for Optical Simulations for Energy Research (BerOSE) of the Helmholtz-Zentrum Berlin für Materialien und Energie GmbH, the Zuse Institute Berlin, and the Freie Universität Berlin. We acknowledge support by the Deutsche Forschungsgemeinschaft within the MATH+ cluster of excellence (EXC-2046/1, project ID: 390685689).

# Author contributions

J.S., K.J., A.S, C.B., M.G. conceived the study. J.S., A.Ö., C.D., M.G. developed the methodology. C.B. chose and interpreted the optical spectrum examples. J.S., K.J., S.B. generated the study data. J.S., A.Ö., M.G. implemented the model and source code. J.S., A.Ö., M.G. conducted the runs. All authors discussed the results. All authors wrote the manuscript. J.S., A.Ö., M.G. designed and produced figures. All authors read and approved the article.

# Competing interests

The authors declare no competing interests.

## Publishers note

Springer Nature remains neutral with regard to jurisdictional claims in published maps and institutional affiliations.

# Supplementary information

## 1.5 Optical Coefficients

The optical behavior of thin films is described by the complex refractive index $\tilde{n}$. It is composed of the refractive index $n$, which governs the phase evolution of the electromagnetic wave in the material, and the extinction coefficient $k$, which describes attenuation due to absorption; both vary with the wavelength $\lambda$ of light.

$$\tilde{n}(\lambda) = n(\lambda) - ik(\lambda) \tag{3}$$

The $n$ and $k$ values used in this work are taken from the `refractiveindex.info` data repository[34]. Incomplete profiles were extrapolated using first- and last-value persistence. Figure S1 visualizes the optical coefficients for all materials. The full data are additionally provided as CSV files in the authors' Hugging Face dataset at Hugging Face.

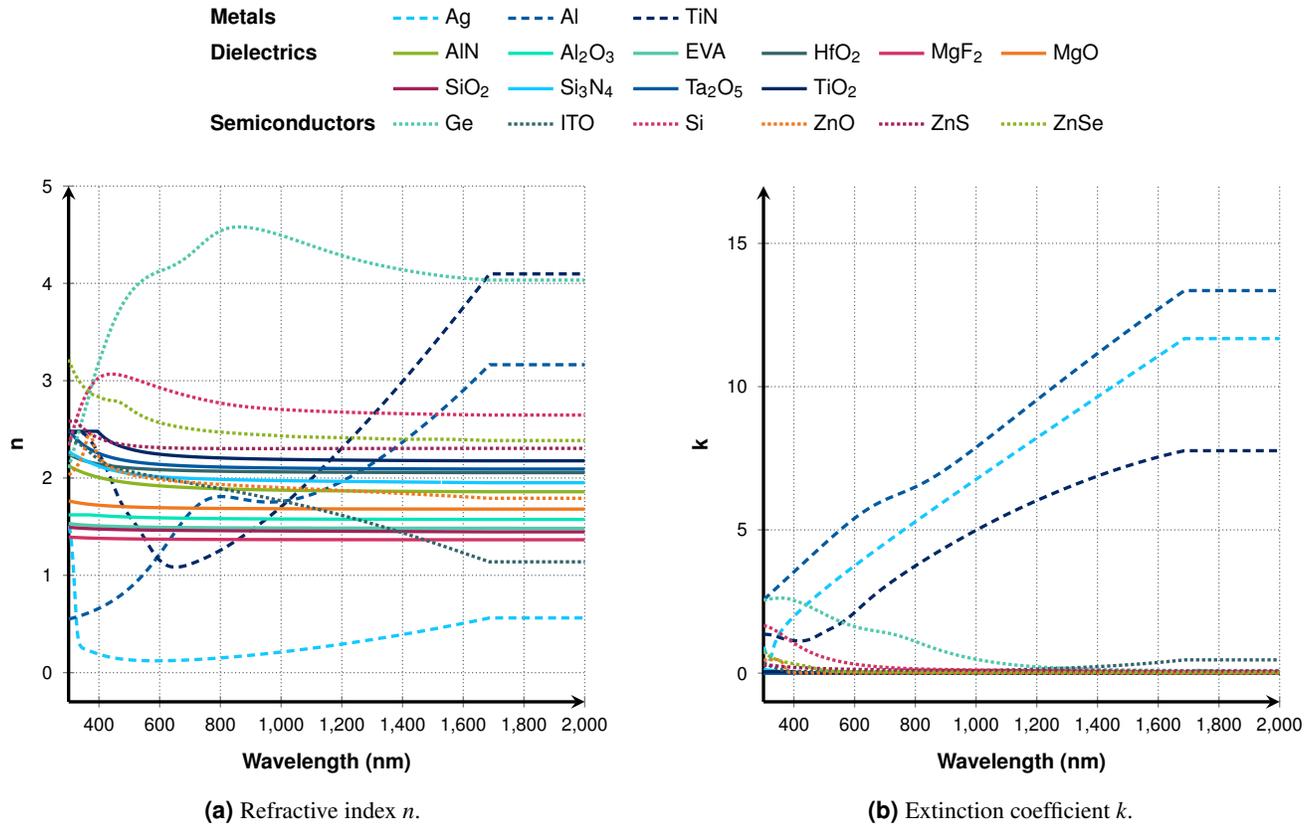

**Figure S1.** Optical coefficients as a function of wavelength for all materials considered in this work. The wavelength range spans 300 nm to 2000 nm.

We additionally clustered the *n*- and *k*-profiles of the materials using hierarchical clustering, with Euclidean distance between the optical-coefficient curves as the similarity metric (cf. Figure S2).

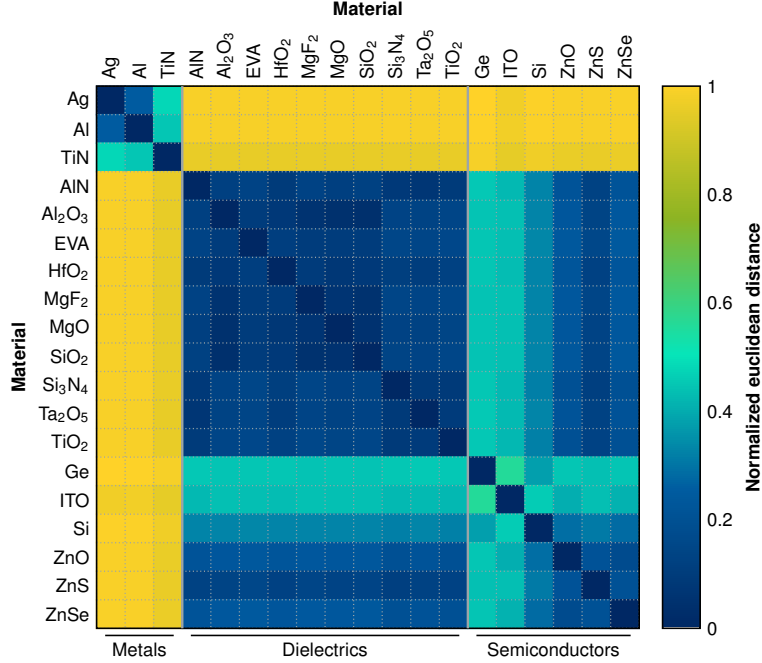

**Figure S2.** Hierarchical clustering of the materials using the normalized Euclidean distance between the complex refractive-index profiles ñ as similarity metric. The distance matrix shows that the major material subclasses, i.e., metals, dielectrics, and semiconductors, are clearly separated.

### 1.6 Computational Environment
We ran the experiments on the distributed-memory, parallel hybrid supercomputer *Lise* at the Zuse Institute Berlin (ZIB). Each of its 42 GPU-accelerated compute nodes is equipped with two 36-core Intel Xeon 8360Y processors at 2.4 GHz base and 3.5 GHz maximum turbo frequency, 1 TB local memory, and one inter-node network card Mellanox InfiniBand NDR interconnect with 400 Gbit/s per port. Each computational node is additionally equipped with four NVIDIA A100-80 GPUs with 80 GB memory connected via NVLink 3.0.

### 1.7 Software Environment
The code used for the experiments was executed in a `Python 3.11.7`[35] compiled with `GNU gcc 11.4.1`. The following libraries and respective versions were used `pandas 2.3.0`[36], `tmm-fast 0.2.1`[27], `torch 2.7.1`[35], `triton 3.3.1`[37], `safetensors 0.6.2`[38] and `CUDA 11.7.64`.

### 1.8 Energy and carbon accounting
We report measured electrical energy consumption for a single training run, 100 inference runs, and the overall development process. From these estimatation, we infer $CO_2$-equivalent emissions and monetary cost.

$CO_2$-equivalent emissions are estimated as

$$m_{CO_2e} = E_{tot} \cdot PUE \cdot I_C, \tag{4}$$

where $E_{tot}$ denotes the measured electrical energy consumption, PUE is the power usage effectiveness of the computing infrastructure, and $I_C$ is the location-specific carbon intensity of electricity generation.

Monetary cost is estimated analogously as

$$C = E_{tot} \cdot PUE \cdot p_C, \tag{5}$$

where $p_C$ denotes the location-specific electricity price per kilowatt-hour.

For the supercomputing system used here, we assume PUE = 1.1. Carbon-intensity values were obtained from Electricity Maps[1] for the selected locations.

---

[1] https://app.electricitymaps.com/map/live/fifteen_minutes

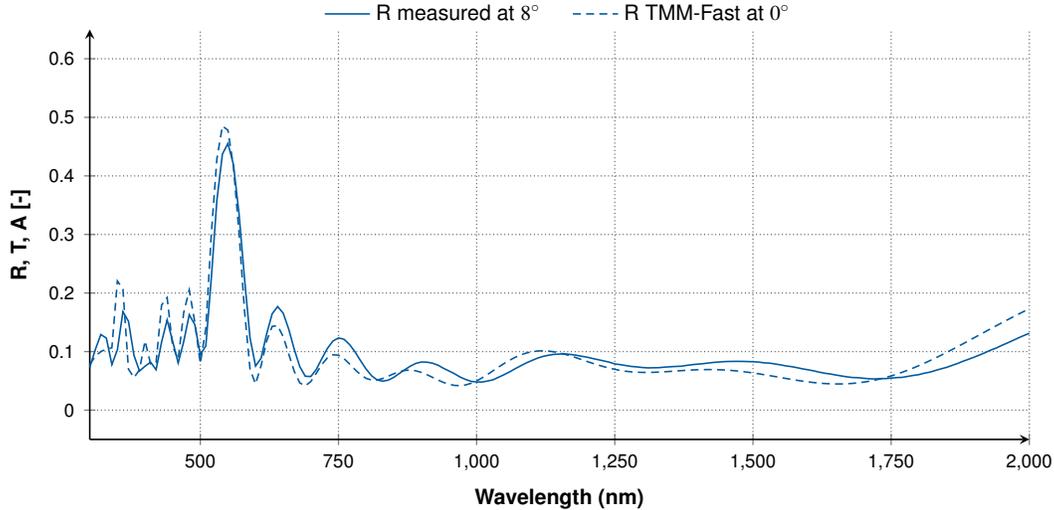

**Figure S3.** Comparison of measured and simulated reflectance for a MorphoColor sample. The measurement is compared with a TMM-Fast simulation under the conditions described in Simulation environment.

### 1.9 Simulation environment

We conducted all simulations with tmm_fast[27], a Python software package designed for vectorized and parallel computation of the optical response of multilayer thin films. It implements the transfer-matrix method (TMM)[39], a standard physics-based approach for calculating the reflection and transmission of light through layered materials. TMM treats each layer as a linear element associated with a $2 \times 2$ matrix that relates the electric and magnetic field amplitudes at the input boundary to those at the output boundary. The overall optical response of the multilayer stack is then obtained by multiplying the individual transfer matrices of all layers.

For all experiments, we assume an incident angle of $0°$, i.e., normal incidence to the thin-film plane. Implemented in PyTorch, tmm_fast can exploit GPUs for accelerated computation and integrates directly with the OptoLlama model implementation.

To assess the realism of the optical model for the MorphoColor reference system, Figure S3 compares the measured reflectance of a textured MorphoColor[22] sample with a simulation obtained using TMM-Fast[27]. The reflectance measurement was performed with an Ulbricht sphere and is therefore limited to an incidence angle of $8°$. For this comparison, the TMM-Fast simulation does not include the texture of the probe. TMM-Fast was evaluated with s-polarized light; this difference is negligible at small incidence angles. Because the MorphoColor filter is deposited on a thick 2.2 mm glass substrate, this layer was treated incoherently.

Overall, the measured and simulated spectra are in good agreement. The remaining small differences in peak position and absolute reflectance level are expected to arise primarily from deviations between the fabricated and modeled structure, in particular from small variations in layer thickness, refractive index mismatches and imperfect layer homogeneity. These effects are not fully captured by the idealized multilayer model, but the comparison nevertheless supports the use of TMM-Fast as an adequate forward model for the design tasks considered in this work.

### 1.10 Hyperparameters

The full set of default hyperparameters used in the experiments is shown in Table S1, unless stated otherwise. All OptoLlama parameters were chosen to match the reproduced OptoGPT baseline as closely as possible and were adjusted only where necessary based on empirical evaluation.

Although OptoLlama contains slightly more parameters than the reproduced OptoGPT baseline (111 555 513 vs. 108 381 113), the relative increase is only about 2.9%. This small difference in model size alone is unlikely to explain the much larger observed improvement in predictive performance. Scaling-law studies have shown that model quality typically improves smoothly with scale, approximately following power-law trends rather than exhibiting abrupt gains from marginal parameter increases. We therefore attribute the performance improvement primarily to the masked diffusion formulation, which denoises the full stack sequence in parallel under spectral conditioning, rather than to parameter count alone.

**Table S1.** Hyperparameters for the reproduced `OptoGPT` baseline and the `OptoLlama` model.

| Hyperparameter | OptoLlama | OptoGPT |
|---|---:|---:|
| Number of processor blocks | 6 | 6 |
| Number of heads per processor block | 8 | 8 |
| Embedding dimension $D$ | 1,024 | 1,024 |
| Positional encoding | Rotary position embedding (RoPE)[40] | Sinusoidal |
| Maximum sequence length | Fixed parametric (here: 20) | Arbitrarily long |
| *Number of parameters* | 111,555,513 | 108,381,113 |
| Optimizer | Adam | Adam |
| Learning rate $\eta$ | $10^{-4}$ | $10^{-4}$ |
| Betas $\beta$ | (0.9, 0.999) | (0.9, 0.999) |
| Batch size $B$ | 256 | 256 |
| Maximum gradient norm for clipping | 1,000 | 1,000 |
| Diffusion steps | 500 | n/a |
| Top-$k$ | 0 | 0 |
| Top-$p$ | 0.9 | 0.9 |
| Temperature | 0.65 | 0.65 |

## 1.11 Training set

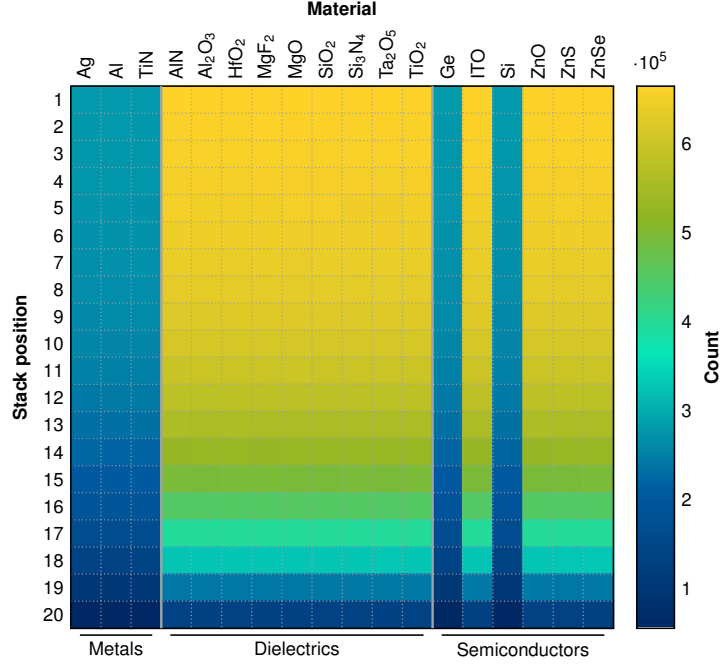

**Figure S4.** Histogram of the $1 \times 10^7$ training samples across material type and stack position. Two broad material groups are visible (Ag, Al, TiN, Ge, Si and AlN, Al$_2$O$_3$, HfO$_2$, MgF$_2$, MgO, SiO$_2$, Si$_3$N$_4$, Ta$_2$O$_5$, TiO$_2$, ITO, ZnO, ZnS, ZnSe). Within each group, the material distribution is approximately uniform across layers. Later stack positions occur less frequently in the training set.

### 1.12 DBR wavelength estimates for the bandstop example

To support the DBR interpretation of the bandstop example discussed in the main text, we estimate the characteristic design wavelengths of the three dominant low/high refractive-index material pairs identified in the predicted stack. For a bilayer consisting of a low-index layer $(n_L, d_L)$ and a high-index layer $(n_H, d_H)$, the first-order DBR wavelength can be approximated as

$$\lambda_B^{(1)} = 2\left(n_L d_L + n_H d_H\right). \tag{6}$$

If the structure operates as a third-order quarter-wave reflector, the corresponding design wavelength is

$$\lambda_B^{(3)} = \frac{\lambda_B^{(1)}}{3}. \tag{7}$$

Using the refractive indices and layer thicknesses extracted for the three material pairs, we obtain:

- MgF$_2$/HfO$_2$ with $n_L = 1.372$, $d_L = 294$ nm, $n_H = 2.093$, $d_H = 200$ nm:

  $$\lambda_B^{(1)} \approx 1644.5 \text{ nm}, \qquad \lambda_B^{(3)} \approx 548.2 \text{ nm}.$$

- MgF$_2$/AlN with $n_L = 1.372$, $d_L = 300$ nm, $n_H = 1.932$, $d_H = 225$ nm:

  $$\lambda_B^{(1)} \approx 1693.2 \text{ nm}, \qquad \lambda_B^{(3)} \approx 564.4 \text{ nm}.$$

- MgF$_2$/TiO$_2$ with $n_L = 1.372$, $d_L = 295$ nm, $n_H = 2.265$, $d_H = 180$ nm:

  $$\lambda_B^{(1)} \approx 1625.5 \text{ nm}, \qquad \lambda_B^{(3)} \approx 541.8 \text{ nm}.$$

These calculations are in excellent agreement with the interpretation given in the main text. In particular, the estimated third-order design wavelengths of approximately 548.2 nm, 564.4 nm, and 541.8 nm closely match the reported values of about

**Table S2.** Estimated first- and third-order wavelengths for the dominant DBR-like substructures in the bandstop example.

| Material pair | $d_L$ [nm] | $d_H$ [nm] | $\lambda_B^{(1)}$ [nm] | $\lambda_B^{(3)}$ [nm] |
|---|---|---|---|---|
| $MgF_2/HfO_2$ | 294 | 200 | 1644.5 | 548.2 |
| $MgF_2/AlN$ | 300 | 225 | 1693.2 | 564.4 |
| $MgF_2/TiO_2$ | 295 | 180 | 1625.5 | 541.8 |

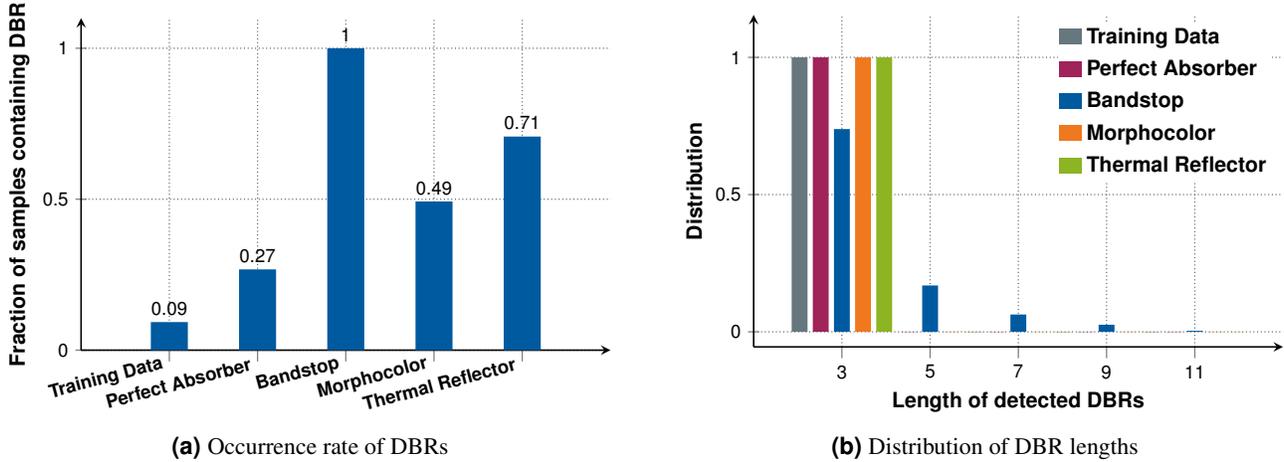

(a) Occurrence rate of DBRs  (b) Distribution of DBR lengths

**Figure S5.** Distributed Bragg Reflector (DBR) Analysis in Training Data and OptoLlama Outputs We have investigated the existence and length of distributed bragg reflectors (DBR) in the data on which OptoLlama was trained as well as on its output given certain targets.

548 nm, 564 nm, and 542 nm, respectively. At the same time, the corresponding first-order DBR wavelengths lie between approximately 1625 nm and 1693 nm, consistent with the statement in the main text that the first-order peaks are located in the 1600 nm to 1700 nm range and therefore outside the spectral region of interest for the bandstop target. Overall, these estimates support the conclusion that the predicted stack contains DBR-like substructures operating near the target bandstop window around 550 nm. Taken together, these estimates support the interpretation that the bandstop prediction is not only spectrally accurate but also structurally aligned with established multilayer design principles.

### 1.13 DBR-like motifs

To assess whether the DBR-like motifs identified in the bandstop example merely reflect memorization, we analyzed their occurrence in both the training data and Monte Carlo samples. Although DBR-like motifs are present in the training set, they are substantially enriched in `OptoLlama` predictions for the bandstop target, suggesting that the model does not simply reproduce training examples but preferentially assembles these motifs under task-specific spectral conditioning. Here, a DBR motif is defined as a contiguous alternating two-material sequence of at least 3 layers, with valid longer motifs occurring only in increments of 2 layers, and with approximate thickness uniformity within each material, such that the maximum thickness difference among all layers of material A is at most 20 nm and the same criterion applies to material B.

Among the 10 million training stacks, 932,237 (9.3%) contained at least one DBR motif. Most detected motifs had length 3 (974,353 occurrences), whereas only 29 motifs of length 5 were found; longer DBR chains did not occur in the training data. For comparison, we analyzed 400 Monte Carlo samples for each of four representative targets. Among these generated stacks, 197 MorphoColor samples, 107 perfect-absorber samples, and 283 thermal-reflector samples contained at least one DBR motif. The bandstop target showed by far the strongest enrichment: all 400 generated samples contained at least one DBR motif, with 1,125 total detected motifs across samples. This target also produced substantially longer DBR chains than those observed in the training set, including motifs of length 7, 9, and 11. Figure S5 summarizes our findings regarding DBR-like motifs. Taken together, these findings support the conclusion that `OptoLlama` does not merely replicate DBR structures present in the training data, but preferentially composes and extends them when required by the target spectrum.

## 1.14 Additional examples

We further illustrate the inverse-design capabilities of `OptoLlama` on two additional target classes shown in Figures S6 and S7: a visually transparent thermal reflector and a perfect absorber. As in the main-text case studies, each example is summarized by four panels showing the reconstructed spectrum, Monte Carlo sampling behavior, wavelength-resolved residuals, and the refractive-index profile of the selected design.

**Thermal reflector.** Figure S6 shows a visually transparent thermal reflector designed to exhibit high transmittance over 300 nm to 1100 nm and high reflectance at longer wavelengths, while remaining non-absorbing across the spectrum. The spectrum in Figure S6a shows that the stack predicted by `OptoLlama` closely matches the target reflectance, absorptance, and transmittance curves, achieving an MAE of 0.036, compared with 0.067 for the template baseline. The corresponding Monte Carlo MAE matrix is shown in Figure S6b, with row-wise best samples indicated by circles and the global best by a star. Consistently, the residuals in Figure S6c remain smaller than those of the template baseline across most of the wavelength range. The refractive-index profile in Figure S6d follows established thin-film design principles: successive layers form a stepwise sinusoidal variation in refractive index at a target design wavelength of 1600 nm, consistent with established broadband reflective-coating motifs while remaining constrained to the materials and thicknesses represented in the training set. This structure enables accurate reproduction of the target response across the ultraviolet, visible, and near-infrared spectral ranges.

**Perfect absorber.** Figure S7 presents a perfect-absorber target over the same wavelength range. The spectrum in Figure S7a shows that, although the template baseline already provides a reasonable approximation (MAE = 0.016), `OptoLlama` further reduces the error to 0.010, yielding a closer match to the idealized target response. The corresponding Monte Carlo MAE matrix is shown in Figure S7b, again with row-wise best samples indicated by circles and the global best by a star. The residuals in Figure S7c confirm the improved reconstruction quality across the spectrum. Finally, the refractive-index profile in Figure S7d shows that the selected design does not reduce to an immediately obvious established design motif, yet still reproduces the target optical response with high fidelity.

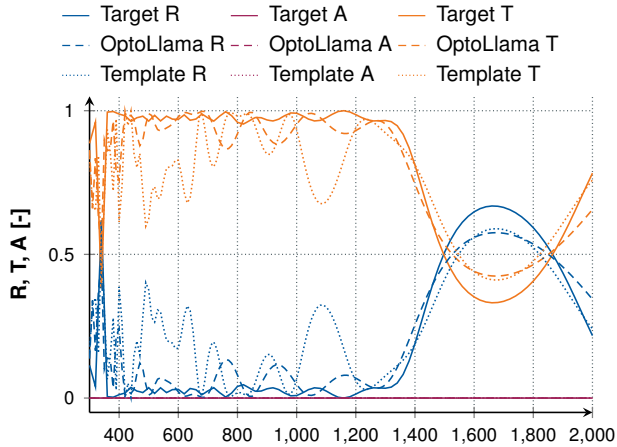

(a) Spectrum for thermal reflector.

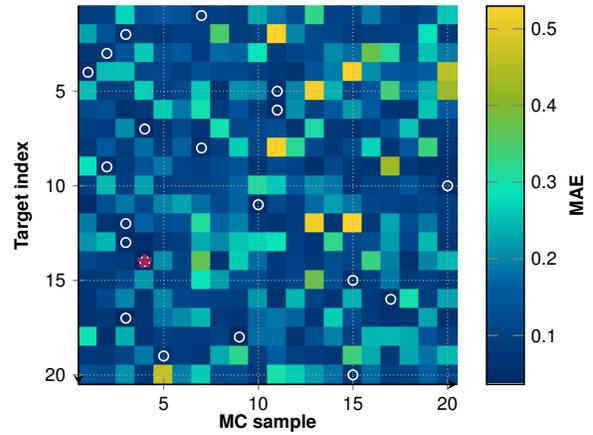

(b) Monte Carlo (MC) sampling for thermal reflector.

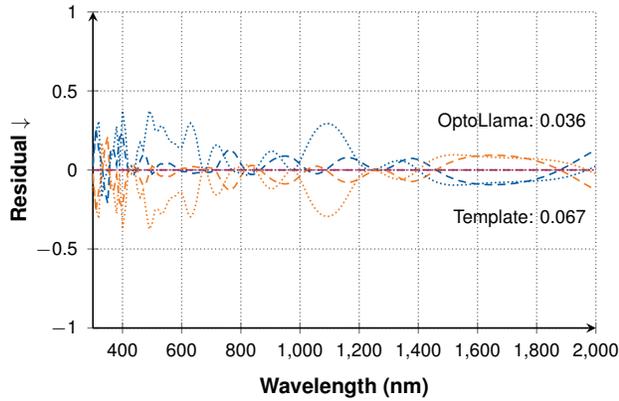

(c) residuals for thermal reflector.

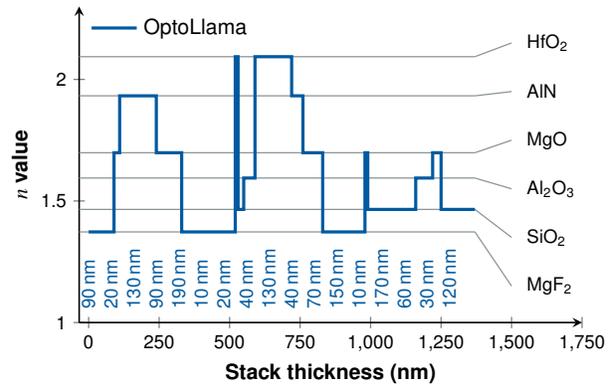

(d) Refractive-index profile of the thermal reflector at 550 nm.

**Figure S6. Example inverse design: visually transparent thermal reflector.** Four panels summarize the target spectrum, stochastic generation process, and best-performing design produced by `OptoLlama`. The top row shows the target and reconstructed reflectance, absorptance, and transmittance spectra (left) and an overview of MAE values across Monte Carlo samples (right), with row-wise best samples and the global best highlighted. The bottom row shows wavelength-resolved residuals (left) and the refractive-index profile of the selected design at 550 nm (right).

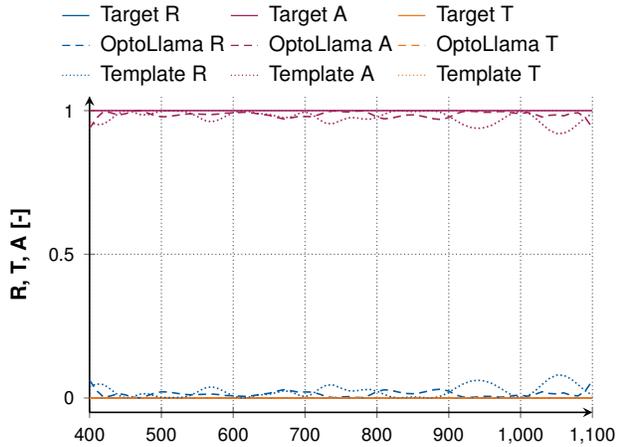

**(a)** Spectrum for perfect absorber.

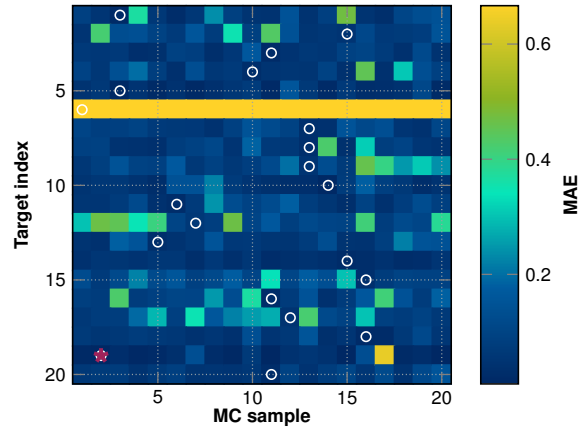

**(b)** Monte Carlo (MC) sampling for perfect absorber.

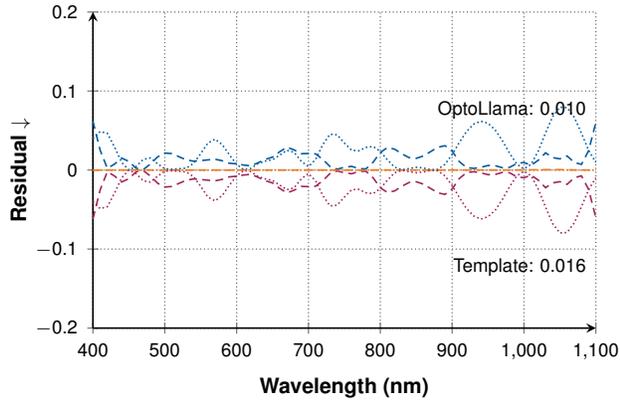

**(c)** residuals for perfect absorber.

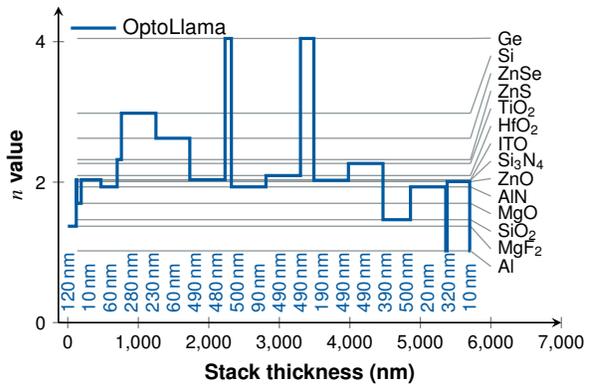

**(d)** Refractive-index profile of the perfect absorber at 550 nm.

**Figure S7. Example inverse design: perfect absorber.** Four panels summarize the target spectrum, stochastic generation process, and best-performing design produced by `OptoLlama`. The top row shows the target and reconstructed reflectance, absorptance, and transmittance spectra (left) and an overview of MAE values across Monte Carlo samples (right), with row-wise best samples and the global best highlighted. The bottom row shows wavelength-resolved residuals (left) and the refractive-index profile of the selected design at 550 nm (right).

## 1.15 Additional variability heatmaps

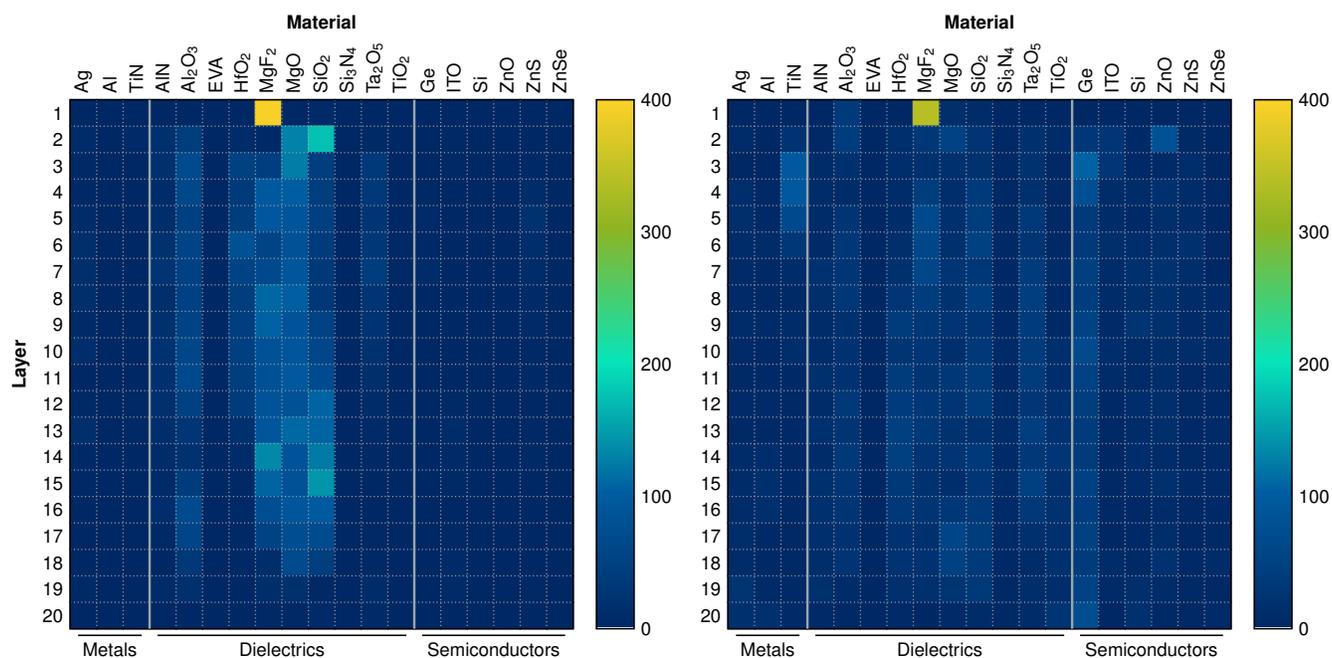

(a) Thermal reflector target.

(b) Perfect absorber target.

**Figure S8. Material variability heatmaps across Monte Carlo samples.** For each target task (a,b), we run `OptoLlama` with $N = 400$ Monte Carlo samples and record the decoded material token at each layer position. Each heatmap shows the resulting material count per thin-film layer (rows) and material class (columns); brighter colors indicate higher frequency. Horizontal brackets group materials into metals, dielectrics, and semiconductors. These distributions summarize model variability and highlight materials that are frequently interchangeable at specific layer positions, as discussed in Variability analysis.

## 1.16 Variability analysis

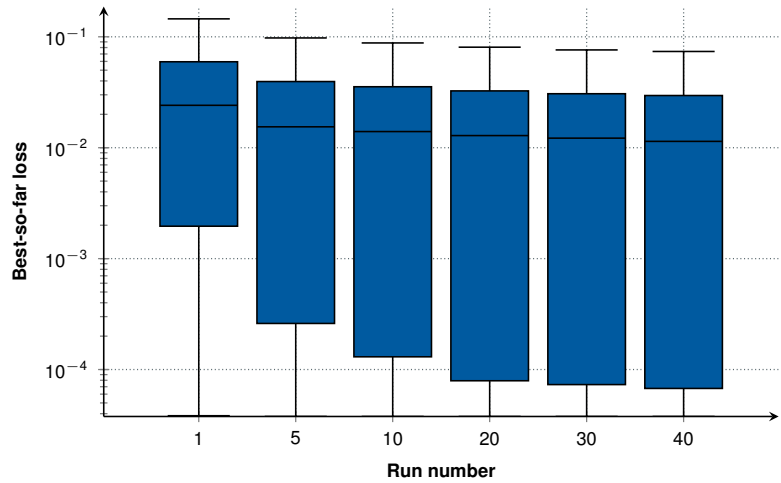

**Figure S9.** Distribution of the best mean absolute error across stochastic runs, aggregated over 3,000 data items.

Since it is possible to evaluate the quality of the predicted stacks, the stack that results in the best loss is chosen out of the multiple Monte Carlo runs that are performed on the model. Consequently, increasing the number of Monte Carlo runs, increases the probability to encounter a prediction that produces an even lower loss. However, performing an infinite number of Monte Carlo runs is not possible due to time and memory constraints. Thus, we have to balance between the benefit of potentially discovering a better predicted stack versus the time and memory cost of performing another run. In order to determine a rule of thumb for the number of Monte Carlo runs one should perform, we investigated the behavior of the best-so-far loss after certain number of runs. Figure S9 shows the evolution of the best-so-far loss on 3000 data items from 1 to 40 runs. While slight improvements continue, the average loss of all data items mostly stagnates at around 10 runs, while some outliers still show improvements at 20 runs.